%% file: main.tex
\documentclass[conference]{IEEEtran}

\usepackage{graphicx}
\usepackage{amsxtra}
\usepackage{amssymb,amsmath}
\usepackage{times,amsmath,epsfig}
\usepackage{cite}
\usepackage{subfigure}
\usepackage{subfloat}
\usepackage[justification=centering]{caption}
\usepackage[noend]{algpseudocode}
\usepackage{times}
\usepackage{boxedminipage}
\usepackage{array,multirow}
\usepackage{tabularx}
\usepackage[linesnumbered,ruled,vlined]{algorithm2e}
\usepackage{pdfsync}
\usepackage{url}
\usepackage[nopatch=eqnum]{microtype}
\usepackage{color}
\usepackage{xpatch}
\usepackage{textcomp}
\xpretocmd{\eqref}{Eq.~}{}{}

\usepackage{comment}

\newcommand{\rememberlines}{\xdef\rememberedlines{\number\value{AlgoLine}}}
\newcommand{\resumenumbering}{\setcounter{AlgoLine}{\rememberedlines}}

\graphicspath{ {images/} }

\newcommand{\remove}[1]{}
\makeatletter
\def\hlinewd#1{%
\noalign{\ifnum0=`}\fi\hrule \@height #1 %
\futurelet\reserved@a\@xhline}
\makeatother

\algnewcommand{\algorithmicgoto}{\textbf{go to}}%
\usepackage{xspace}
\algnewcommand{\Goto}{\algorithmicgoto\xspace}%
\algnewcommand{\Label}{\State\unskip}

\newcommand\blfootnote[1]{%
  \begingroup
  \renewcommand\thefootnote{}\footnote{#1}%
  \addtocounter{footnote}{-1}%
  \endgroup
}

\IEEEoverridecommandlockouts % override IEEE command. this will enable \thanks{} which is disabled by IEEE style

\def\BibTeX{{\rm B\kern-.05em{\sc i\kern-.025em b}\kern-.08em
    T\kern-.1667em\lower.7ex\hbox{E}\kern-.125emX}}

\begin{document}

%----------------------------------------------------------------------
% Title Information, Abstract and Keywords
%----------------------------------------------------------------------
%\title{An Efficient Admission Control and Scheduling Algorithm for IEEE 802.11ad Traffic}
%\title{An Efficient Admission Control and Scheduling Algorithm for Isochronous and Asynchronous IEEE 802.11ad Traffic}
\title{An Admission Control Algorithm for Isochronous and Asynchronous Traffic in IEEE 802.11ad MAC}
\newcommand*\samethanks[1][\value{footnote}]{\footnotemark[#1]}
% Author names and affiliations
%\author{\IEEEauthorblockN{}

\author{\IEEEauthorblockN{
Anirudha Sahoo} 
 Communications Technology Laboratory, \\
 \IEEEauthorblockA{National Institute of Standards and Technology, Gaithersburg, Maryland, USA \\
 Email: anirud@nist.gov}}

\maketitle
%\pagenumbering{arabic}
%\pagestyle{plain}

%\pagenumbering{arabic}
%\pagestyle{plain}

\begin{abstract}
\input{abstract}

\end{abstract}


%\keywords{IEEE 802.11ad, MAC, admission control, scheduling}

\maketitle

%----------------------------------------------------------------------
% SECTION I:
%----------------------------------------------------------------------
\section{Introduction}
\label{sec-introduction}
\input{introduction}

%----------------------------------------------------------------------
% SECTION II:
%----------------------------------------------------------------------

%----------------------------------------------------------------------
% SECTION III:
%----------------------------------------------------------------------
%\section{Methodology}
%\label{sec-approach}
%\input{problem}
%\input{algo}

%----------------------------------------------------------------------
% SECTION IV:
%----------------------------------------------------------------------
%\section{Admission Control and Scheduling of Isochronous and Asynchronous Traffic}
\section{A Primer on IEEE 802.11ad MAC}
\label{sec-overview}
\input{overview}

%----------------------------------------------------------------------
% SECTION V:
%----------------------------------------------------------------------
%\section{Admission Control and Scheduling}
%\input{scheduler}

%\section{Simulation Results}
%\input{results.tex}

%----------------------------------------------------------------------
% SECTION VI:
%----------------------------------------------------------------------
\section{Related Work}
\label{sec:related_work}
\input{relatedwork}

\section{Conclusion}
\label{sec-conclusion}
\input{conclusion}

\vspace{-0.1in}
\bibliographystyle{IEEEtran}

\bibliography{master}

\end{document}

%% file: abstract.tex
Due to availability of large amount of bandwidth in the 60~GHz band and support of contention-free
channel access  called \textit{Service Period} (SP),  the IEEE 802.11ad/ay Wi-Fi standard is well
suited for low latency and high data rate applications.
IEEE 802.11ad supports two types of SP user
traffic: \textit{isochronous} and \textit{asynchronous}.
These user traffic
need guaranteed SP duration before their
respective deadlines. Hence, admission control plays an important role
in an IEEE 802.11ad system. In an earlier work, we
studied admission control and scheduling of isochronous and
asynchronous traffic in 
an IEEE 802.11ad system, but we assumed the
asynchronous requests to be periodic to keep the
algorithm simple. That assumption resulted
in overallocation of resource and potential degradation
of performance. 
In this paper, we present an admission control
algorithm  which does not make
such assumption and yet still maintains a linear run time complexity and allocates resources to
the requests in a proportional fair manner. We
provide arguments to establish correctness of
the algorithm in terms of guaranteeing SP
allocation to the requests before their respective
deadlines.

%% file: introduction.tex
A large amount of unlicensed bandwidth is available
in the millimeter wave (mmWave) 60~GHz band. In recent times, the use of high
data rate and low latency applications such as
8K TV, Virtual Reality (VR) and Augmented Reality (AR) are on the rise.  These factors
have led to development of
standards for the next generation Wi-Fi
such as IEEE 802.11ad and its successor IEEE 802.11ay. In terms of providing high data rate and
low latency, the IEEE 802.11ad Medium Access Control (MAC) plays an important role. 
The IEEE 802.11ad MAC supports contention-free channel access
between a pair of stations,
called \textit{service period} (SP).
It handles two types of SP traffic: \textit{isochronous} and
\textit{asynchronous}. Isochronous traffic needs
guaranteed allocation of certain minimum 
\textit{service period} (SP) duration in every period.
Asynchronous
traffic, on the other hand, requires one time
guaranteed allocation of SP before its deadline. 
So, every admitted isochronous and asynchronous request must
be guaranteed its requested SP duration before its
deadline. These two types of traffic are requested
using 
\textit{Add Traffic Stream} (ADDTS) request\footnote{Although asynchronous traffic can be requested by other means, in this paper, we only consider ADDTS based method.}, which carries the
traffic parameters~\cite{802.11ad_standard}. Due
to the stringent allocation and deadline requirements, admission control and scheduling of
requests is an important component of an IEEE 802.11ad system.
We have studied  admission control and scheduling
algorithms for only isochronous requests 
in an IEEE 802.11ad system~\cite{mswim2021, tmc2022}. Further, in~\cite{vtc2023}, we presented admission control and scheduling of both
isochronous and asynchronous traffic. However, in that work, we treat asynchronous traffic as periodic
traffic, which allows us to use the simple
admissibility criteria used in 
\textit{Earliest Deadline
First} (EDF) scheduling for 
periodic Central Processing Unit (CPU) tasks
in a realtime system~\cite{liu73}. However,
this assumption results in overallocation of
resources to asynchronous requests, leading to
potential performance loss in terms of admitting
fewer requests. In this paper, we present an admission control algorithm called, Efficient Admission Control
for Isochronous and Asynchronous Requests (EACIAR),
which does not make such assumption and yet still
maintains a linear run time complexity as well as
allocates resources in a proportional fair manner.
Periodic and aperiodic CPU tasks are quite similar to
isochronous and asynchronous traffic respectively.
Admission control and scheduling of periodic
and aperiodic CPU tasks have been studied
in the literature~\cite{liu73, jeffay91, thuel94, chetto89}. But some of them have severe limitations, whereas some others cannot be
directly applied to IEEE 802.11ad MAC due to
difference in traffic description (please
refer to Section~\ref{sec:related_work} for more
detailed discussion). We present the detailed
admission control algorithm along with the
associated scheduling. We provide arguments 
to establish correctness of the algorithm in the sense
that allocation requirement of every admitted request
is met before its deadline. To the best of our knowledge, this is the first comprehensive
admission control and scheduling algorithm for
IEEE 802.11ad MAC without any restrictions on
the request type or on the traffic parameter values of
the requests\blfootnote{U.S. Government work, not subject to U.S. Copyright.}.

\begin{comment}
We present detailed performance results of ACIAR algorithm with respect to different performance metrics. Especially, we investigate the
impact of asynchronous traffic on the performance
of isochronous traffic.
Our results show that the
presence of asynchronous requests  cause
the system to admit lesser number of
isochronous requests and the admitted
isochronous requests achieve lower channel
utilization. However, the performance
of those admitted isochronous requests in terms of 
other metrics such as amount of channel time
allocated to each request and delay is better when
they share the system with asynchronous requests. 
Our simulation results on overallocation show that the ACIAR algorithm may incur performance
loss when the system runs with a very high
isochronous request load.

The main contributions of this work are: i) we present 
admission control and scheduling algorithms that can handle both
isochronous and asynchronous requests in an IEEE 802.11ad system, ii) we provide a detailed
analysis of performance loss due to overallocation 
of resources to asynchronous request since they
are treated 
as periodic requests in our SAC algorithm and the condition when
the performance loss may occur (which will be useful in a real system with hybrid admission control scheme), iii) we present detailed simulation results of
our SAC algorithm in different scenarios and configurations
to show the effect of presence of asynchronous requests
on the isochronous requests.
\end{comment}

%% file: overview.tex
\subsection{IEEE 802.11ad Medium Access}
The medium access duration in
IEEE 802.11ad is divided into 
an infinite sequence of time intervals, called
\textit{Beacon Interval} (BI). 
The length of a BI duration is specified in Time Units
(TU), where $\mathrm{1\,TU=1024\, \mu}$s. In this paper, we
represent a BI duration as a sequence of 1\,$\mu$s time slots. A BI has two parts:
a Beacon Header Interval (BHI) followed by a 
Data Transmission Interval (DTI). The DTI is 
primarily used for data transmission among 
IEEE 802.11ad
Stations (STAs) and Personal Basic Service Set (PBSS) Control Point/Access Point (PCP/AP). An IEEE 802.11ad
station can access the channel in a DTI using 
Contention Based Access Period (CBAP) or 
Service Period (SP) mechanism. When using
CBAP, a station uses a contention based scheme called Enhanced Distributed Channel Access (EDCA)~\cite{802.11ad_standard}. An SP, on the other hand,
is used between two stations or between a station and
its PCP/AP to have a contention free channel access. 
Before a DTI period starts, the
schedules of CBAP and SP in a DTI are broadcast by the
PCP/AP to its stations in a Directional Multi Gigabit (DMG) Beacon frame in the Beacon Transmission Interval (BTI) or in the Announce frame in the Announcement Transmission Interval (ATI) of the BHI~\cite{802.11ad_standard}.

\subsection{IEEE802.11ad Traffic Parameters}
The Traffic Specification (TSpec) element in the ADDTS request carries the traffic parameters for which resources need to be allocated. For isochronous requests, channel time allocation duration
is repeated in every
period, whereas for asynchronous traffic, allocation is one time.
The main traffic parameters (used in the DMG TSpec frame) of isochronous traffic are~\cite{802.11ad_standard, mswim2021}:
\begin{itemize}
    \item Allocation Period ($P$): Period over which allocation repeats. It can be an integer multiple or integer fraction of a BI.
    \item Minimum Allocation ($C_{min}$): Minimum acceptable allocation in microseconds in each allocation period. If the request is accepted, the
    PCP/AP must guarantee at least this duration to the STA in every allocation 
    period.
     \item Maximum Allocation ($C_{max}$): Requested allocation in microseconds in each allocation period. This is the maximum
     duration that can be allocated to the user in each allocation
     period.
\begin{comment}
     \item Minimum Duration: Minimum duration in microseconds in each allocation period. An allocation may be split into multiple fragments. Each fragment must be larger than  or equal to this duration. The user can set this value to zero to indicate that this parameter should not be considered. In this study, we assume this parameter to be zero to keep the problem simple.
\end{comment}
\end{itemize}
Asynchronous requests also use the same DMG TSpec parameters. However, 
the minimum allocation is one time allocation that must be done before the allocation period (or \textit{deadline}).
Maximum allocation field is reserved.  

\section{Admission Control}

\subsection{Isochronous and Asynchronous Traffic Model}
A request $T_i$ is an isochronous (resp. asynchronous) request
when $T_i.reqType$ is \textit{ISO} (resp. \textit{ASYNC}). For the ease of notation, we
will use those two terms throughout this paper
to refer to the respective requests. In addition,
we will use \textit{ISO\textsubscript{M}} and \textit{ISO\textsubscript{F}} for ISO requests whose
periods are integer multiple of a BI and integer
fraction of a BI, respectively.
The period, minimum allocation, and maximum
allocation of an isochronous request $T_i$ are denoted by
$T_i.P$, $T_i.C_{min}$ and $T_i.C_{max}$, respectively. The actual
(or operational) channel time allocation to isochronous request $T_i$ is 
denoted as $T_i.C_{op}$ and should satisfy 
$T_i.C_{min} \le T_i.C_{op} \le T_i.C_{max}$. When
there are no ASYNC requests in the system, an ISO request $T_i$ would be allocated $T_i.C_{op}$. However, when there is at least one ASYNC request
in the system, the operational channel time of
an ISO request is decided based on
the schedule carried in \textit{long\_schedule[]} (see Algorithm~\ref{alg:adm_control} for its computation). 
The period (or deadline) and
minimum allocation of asynchronous request $T_i$ are denoted by
$T_i.P$ and $T_i.C_{min}$, respectively. The operational
allocation $T_i.C_{op}$, in this case, is equal to $T_i.C_{min}$. An isochronous request has allocation
request in every period. Such allocation requests are
referred to as \textit{jobs} of the request. Since an \textit{ISO\textsubscript{F}} request has multiple periods in a BI, it will have multiple jobs in a BI. An
$ISO_M$ request, on the other hand, will have one job in multiple BIs.
Since
an asynchronous request does not
repeat beyond its deadline, it has exactly one job in
its lifetime. A job of a request becomes ready to
be allocated at the beginning of its period, which
we refer to as the \textit{release time} of the job.
A job
of a request may be allocated one contiguous block or it may be
broken into 
multiple \textit{fragments} to fit smaller idle durations in 
a BI.

\subsection{CPU Scheduling of Periodic Tasks}
\label{sec:cpu_schedule}
ISO requests are periodic and hence, have similarity
with periodic tasks in CPU scheduling which has been studied in the literature~\cite{liu73, jeffay91, thuel94, chetto89}. So, we design scheduling of
ISO request based on theories developed for CPU scheduling of periodic tasks. A periodic task in CPU scheduling has two parameters
\textit{($C_i$, $P_i$)}, where $C_i$ is the duration of the task and $P_i$ is the period as well as the deadline of the task~\cite{vtc2023}.  
The feasibility or admissibility of a set of $n$ preemptive periodic tasks for an EDF scheduler is given by~\cite{liu73}
\begin{eqnarray}
\sum_{i=1}^n \frac{C_i}{P_i} \le 1.  \label{eq:util}
\end{eqnarray}

%\subsection{Admission Control of Isochronous and Asynchronous Requests}

\subsection{Admission Control at a High Level}
\begin{figure}[!t]
\centering
\includegraphics[scale=0.35] {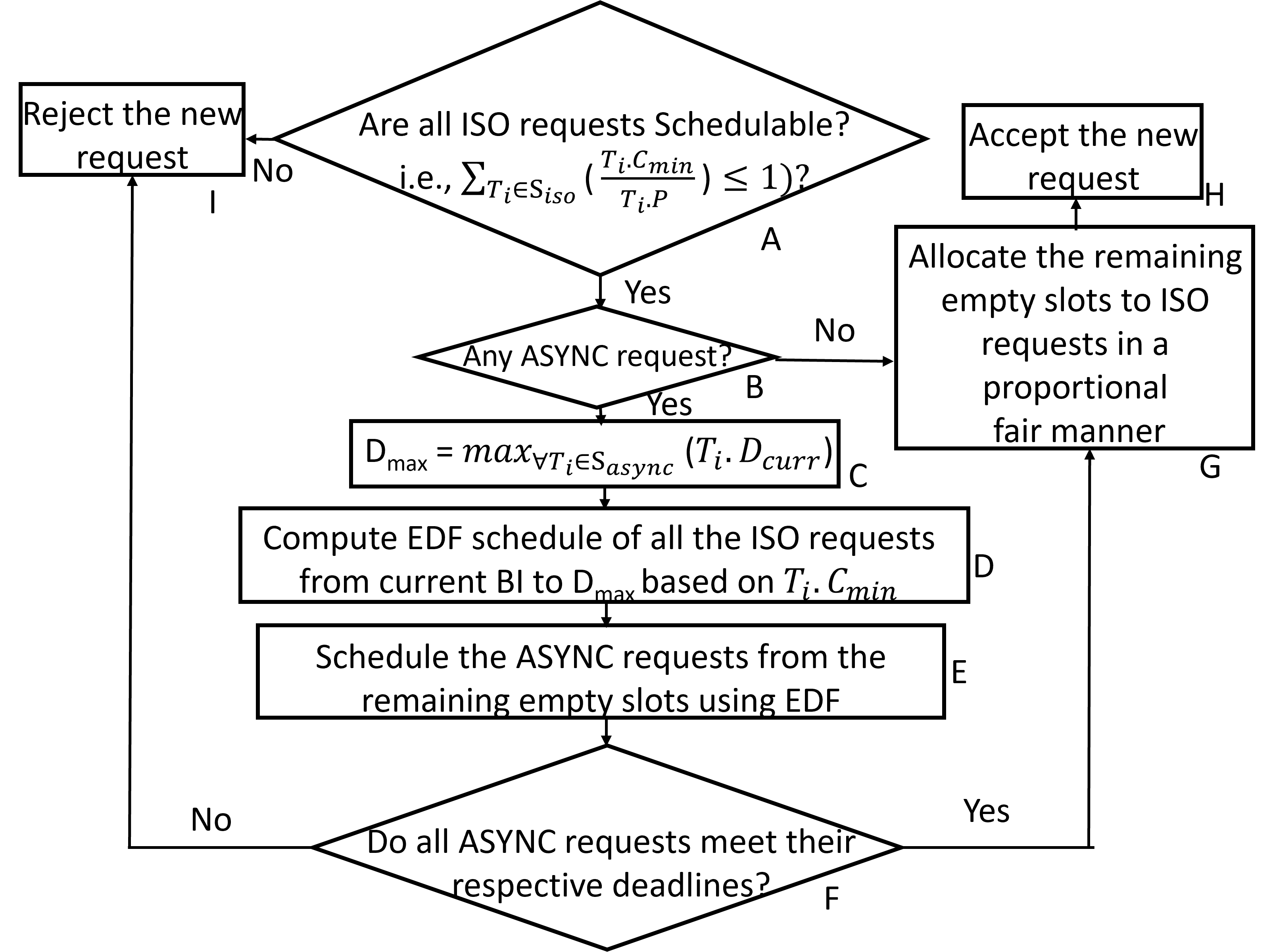}
%\includegraphics[scale=0.3]{figures/perf_loss.png}
%\caption{Illustration of Condition when Overallocation may Lead to Performance Loss}
\caption{Flowchart of Admission Control at a High Level}
\label{fig:admission_control_high_level}
\end{figure}

The overall admission control flowchart is shown in Fig.~\ref{fig:admission_control_high_level}. The basic idea is to
check whether each request (both existing and the
new request) is schedulable before its deadline, when a new request arrives. 
It first checks if all the ISO requests are schedulable or not based on their respective $C_{min}$ demand (Decision Box A). Notice that the Decision Box A utilizes admissibility criteria of periodic CPU tasks presented in~\eqref{eq:util} based on $T_i.C_{min}$.  If all the ISO requests are not schedulable, then the new
request is rejected (Box I). Otherwise, it checks if there is any ASYNC request in the system (Decision Box B). If not, then it allocates the remaining empty slots to the ISO requests in a
proportional fair manner (Box G) and accepts
the new request (Box H). Otherwise, it computes
$D_{max}$, which is the maximum over all the current
deadlines of ASYNC requests (Box C). Then it computes EDF schedule of all the ISO requests from current BI to $D_{max}$ (Box D). ASYNC requests are then scheduled from the remaining empty slots using EDF (Box E). If all the ASYNC requests meet their respective deadlines (decision Box F), then any remaining empty slots are allocated
to ISO requests in a proportional fair manner (Box G) and the new request is accepted (Box H). Otherwise the new request is rejected (Box I).

\begin{comment}
It follows EDF with schedule \textit{as late as possible}. This approach (as opposed to EDF with schedule \textit{as early as possible}) could potentially leave more empty slots before the deadlines of ASYNC requests.

The TSpec of isochronous traffic has a range of duration, unlike the CPU scheduling of a task which has a single duration.
So, an admission control algorithm not only determines whether
a new request can be admitted or not, but also computes  $C_{op}$. We are able
to use the admissibility criteria \eqref{eq:util} by treating asynchronous requests as periodic.

Let us assume that there are $(n-1)$ requests, running at their respective $C^{op}$s, already in the system.
The newly arriving $n^{th}$ request, $T_n$, which could be an
isochronous or asynchronous request, has the operational allocation
$C^{op}_n$. Then $T_n$ is admitted if and only if

\begin{eqnarray}
U + \frac{C^{op}_n}{P_n} \le 1,  \label{eq:util1}
\end{eqnarray}
where $U = \sum_{i=1}^{(n-1)} \frac{C^{op}_i}{P_i}$ is the utilization of the system due to already
admitted requests. 
\end{comment}

\subsection{Maintenance of Information and Operation in every BI}
For admission control and scheduling
purposes, a table for each admitted 
request is maintained. The content of
the table is shown in Table~\ref{tab:req_tbl}. In addition, two global variables, $\mathcal{S}_{iso}$  and $\mathcal{S}_{async}$ which carries the set of existing
ISO and ASYNC requests respectively, are maintained.
Once
the schedule for the current BI is computed and executed, the parameters of the table are
updated as follows:
\begin{itemize}
    \item For each \textit{ISO\textsubscript{M}} request and each ASYNC request, $C_{remain}$ is decremented by the duration (in $\mu s$) allocated to the request in the current BI. Note that, for an \textit{ISO\textsubscript{M}} request, its $C_{remain}$ can become less than zero (which means the request has been allocated more than its $C_{min}$), since it may be allocated more than $C_{min}$ in a given period. \textit{ISO\textsubscript{F}} requests are fully allocated within a BI and hence, $C_{remain}$ is not relevant for them.

    \item For all requests, $t_{remain\_life}$ is decremented by one. If it becomes zero, it
    signifies that the request is finished. Hence,
    that request is removed from the set $\mathcal{S}_{iso}$ or $\mathcal{S}_{async}$ depending on its reqType and
    the corresponding table entry is removed.
    At this point, if there are no
    ASYNC requests, then an EDF schedule is followed. Otherwise, admission control algorithm is executed again to recalculate \textit{long\_schedule[]} which is followed from the next BI.

    \item For each \textit{ISO\textsubscript{M}} request and  ASYNC request, its $D_{curr}$ is decremented by one. For an \textit{ISO\textsubscript{M}} request, if its $D_{curr}$ becomes zero and $t_{remain\_life}$ is greater than zero, then $C_{remain}$ is reset to $C_{min}$ and $D_{curr}$ is reset to $P$. Note that for AYSNC requests when $D_{curr}$ becomes zero, $t_{remain\_life}$ should also become zero.
\end{itemize}

\begin{table}[]
\centering
\caption{Request Table Entry}
\label{tab:req_tbl}
%\begin{tabular}{|c|c|c|}
\begin{tabular}{|p{0.2\columnwidth}|p{0.1\columnwidth}|p{0.55\columnwidth}|}
\hline
\textbf{Parameter}   & \textbf{Unit} & \textbf{Description} \\ \hline
  reqType &        'ISO' or 'ASYNC' &  set to 'ISO' for isochronous request or 'ASYNC for asynchronous requests\\ \hline

  $C_{min}$, $C_{max}$ & $\mu s$ & $C_{min}$ and $C_{max}$ of the request. $C_{max}$ is not relevant for ASYNC requests \\ \hline

  P  &  BI  & period of Request, if \textit{reqType} is ISO. Original deadline of the request if \textit{reqType} is ASYNC. \\ \hline

  $C_{op}$  & $\mu s$ & operational channel time allocation. Only relevant for ISO requests. \\ \hline

  $C_{remain}$  & $\mu s$ & remaining channel time allocation still needed  before its current deadline to satisfy $C_{min}$. Only relevant for ASYNC requests and \textit{ISO\textsubscript{M}} requests. \\ \hline

  $t_{remain\_life}$ & BI & time remaining for this request to leave the system \\ \hline
  
  $D_{curr}$  &  BI  & current deadline of the request. Only relevant for ASYNC requests and \textit{ISO\textsubscript{M}} requests. \\ \hline

\hline
\end{tabular}
\end{table}

\begin{algorithm*}
\small
%\footnotesize
%\scriptsize
\LinesNumbered
% This is to hide end and get the last vertical line straight
\SetKwBlock{Begin}{Begin}{}
\SetAlgoLined

\caption{\textbf{Efficient Admission Control for Isochronous and Asynchronous Requests (EACIAR)}} \label{alg:adm_control}
%\begin{algorithmic}[1]
\KwIn{the new request $T_n$, the set of existing isochronous requests $\mathcal{S}_{iso}$ and the set of existing asynchronous requests $\mathcal{S}_{async}$}
%\State

\KwOut{ACCEPT or REJECT; if no ASYNC request present and the new request is accepted, $C_{op}$ of every ISO request; if ASYNC request present and the new request is accepted, $long\_schedule[]$}

\Begin{
% This is to restore vline mode
  \SetAlgoVlined
  
\lIf {($T_n.reqType$ == ISO)} { 
 $\mathcal{S}_{iso} = \mathcal{S}_{iso} \cup \{T_n\}$
}
\lElse {
$\mathcal{S}_{async} = \mathcal{S}_{async} \cup \{T_n\}$
}

$U_{min}^n$ = $\Sigma_{T_i \in \mathcal{S}_{iso}} (\frac{T_i \cdot C_{min}}{T_i \cdot P})$ \;

\lIf {($U_{min}^n> 1$)} { \label{alg:iso_check}
  \Goto \texttt{RJ }
}

\If (\tcp*[h]{no ASYNC requests and all ISO requests are schedulable}) {($\mathcal{S}_{async} == \emptyset$)}  { \label{alg:iso_alloc1}
  %\If {($\Sigma_{T_i \in \mathcal{S}_{iso}}) (\frac{T_i \cdot C_{min}}{T_i \cdot P}) \le 1)$} {
  $U_{surplus} = 1-U_{min}^n$ \;
  $\Delta u_{tot}$ =  $\Sigma_{T_i \in \mathcal{S}_{iso}} (\frac{T_i.C_{max} - T_i.C_{min}}{T_i.P})$ \;
  \For {each $T_i \in \mathcal{S}_{iso}$} {
     $T_i.C_{op} = T_i.C_{min} + \min{(1, \frac{U_{surplus}}{\Delta u_{tot}})} \cdot (T_i.C_{max} - T_i.C_{min})$;  \tcp{proportional fair allocation of surplus}
  }
    return ACCEPT, (for \ each \ $T_i \in \mathcal{S}_{iso} \ (T_i.C_{op}))$ \;
  } \label{alg:iso_alloc2}
  %\Else \, \Goto{RJ}
  %\Else {\Goto \texttt{RJ} \;}

$\mathcal{S}^{temp}_{iso} = \mathcal{S}_{iso}$ ;
$D_{max} = \max_{T_i \in \mathcal{S}_{async}} (T_i. D_{curr})$ \;
long\_sched[$D_{max}$ * BI] = 0 \;
\For {i = 1 to $D_{max}$} { \label{alg:for}
  $Schedule[1 \ldots BI] = 0$ \;
  \tcp*[h]{process ISO requests whose periods are integer fraction of a BI} \\
  $J^{iso}_{frac}$ = \{array of jobs of ISO requests in $\mathcal{S}^{temp}_{iso}$ in the current BI whose periods are integer fraction of a BI\} \;
  $J^{iso}_{sort\_frac}$ = \{sorted array of jobs in $J^{iso}_{frac}$ in a non-decreasing order of their current deadlines\} \;
  k = 0 \;
  \While {(k $<$ len($J^{iso}_{sort\_frac}$)} {
     R = $J^{iso}_{sort\_frac}[k].release$;
     D = $J^{iso}_{sort\_frac}[k].P$;
     $C_{min}$ = $J^{iso}_{sort\_frac}[k].C_{min}$ \;
     \While {($C_{min} > 0 $)} { \label{alg:iso_frac_alloc}
        % this block of code is not needed. The else part covers this part also.
        %\If{$Schedule[R \ldots (R+C_{min}] == 0$} { \tcc{contiguous $C_{min}$ slots available}
           %$Schedule[R \ldots (R+C_{min}] = 1$ \;
           %break \;
         %}

         %\Else {
            $avail$ = number of contiguous slots available from position 'R' in Schedule[] \;
            $avail$ = $\min(avail, C_{min})$ \;
            $Schedule[R \ldots (R+avail] = 1$ \;
            $C_{min} -= avail$ \;
            R = next empty slot position in Schedule[] \;
            \lIf {($R > D$)} { 
              print error and exit; \tcp*[h]{this should not happen, since we have done the utilization test for ISO requests}
            }
        %}
     }
     $k^{++}$ \;
  }
  
    \tcp*[h]{process ISO requests whose periods are integer multiple of a BI} \\
  $J^{iso}_{mult}$ = \{array of jobs of ISO requests in $\mathcal{S}^{temp}_{iso}$ in the current BI whose periods are integer multiple of a BI\} \;
  $J^{iso}_{sort\_mult}$ = \{sorted array of jobs in $J^{iso}_{mult}$ in a non-decreasing order of their current deadlines\} \;
  k = 0 \;
  \While {(k $<$ len($J^{iso}_{sort\_mult}$)} {
     R = 1 ;
     D = $J^{iso}_{sort\_mult}[k].D_{curr}$ ;
     $C_{remain}$ = $J^{iso}_{sort\_mult}[k].C_{remain}$ \;
     \While {($C_{remain} > 0 $)} { \label{alg:iso_int_alloc}
            $avail$ = number of contiguous slots available from position 'R' in Schedule[] \;
            $avail$ = $\min(avail, C_{remain})$ \;
            $Schedule[R \ldots (R+avail] = 1$ \;
            $C_{remain} -= avail$ \;
            R = next empty slot position in Schedule[] or (-1) if Schedule[] is full (all 1)\;
            \lIf {($R == -1$)} { 
              print error and exit; \tcp*[h]{this should not happen, since we have done the utilization test for ISO requests} 
            }
     }
     $k^{++}$ \;
  }
  %update $C_{remain}$, $D_{curr}$ of ISO requests whose $P= n \cdot BI$ \;
  long\_schedule[i] = schedule[] \;
  \For {each $T_i \in \mathcal{S}^{temp}_{iso}$} {
     \lIf (\tcp*[h]{Request $T_i$ leaves the system}) {($T_i.t_{remaining\_life} == i$)} {
     $\mathcal{S}^{temp}_{iso}$ = $\mathcal{S}^{temp}_{iso} \setminus \{T_i\}$
     }
  }
  update current BI to the next BI \;
  }
  tot\_async\_dur = 0 \;
\For {each $T_i \in \mathcal{S}_{async}$} { \label{alg:tot_async_dur}
  tot\_async\_dur += $T_i.C_{remain}$ \; 
}
num\_empty\_slots = number of empty slots in long\_schedule[] \;
\If (\tcp*[h]{not enough empty slots to schedule all the ASYNC requests}) {(tot\_async\_dur $>$ num\_empty\_slots)} {
 Goto \texttt{RJ} \; 
}
} % of Begin
\rememberlines %remember line number

\end{algorithm*}

\begin{algorithm*}
\small
%\footnotesize
%\scriptsize
\LinesNumbered
\resumenumbering % resume from previous block
%This is to hide Begin keyword
\SetKwBlock{Begin}{}{end}
\Begin{

\tcp*[h]{process ASYNC requests} \\
$\mathcal{S}^{async}_{sort}$ = \{sorted array of ASYNC requests in a non-decreasing order of their current deadlines\} \;
k = 0; tot\_c\_async = 0 \;
\While {(k $<$ len($S^{aysnc}_{sort}$)} {
     R = 1 \;
     D = $\mathcal{S}^{async}_{sort}[k].D_{curr}$ \;
     $C_{remain}$ = $\mathcal{S}^{async}_{sort}[k].C_{remain}$ \;
     \While {($C_{remain} > 0 $)} { \label{alg:async_alloc}
            $avail$ = number of contiguous slots available from position 'R' in Schedule[] \;
            $avail$ = $\min(avail, C_{remain})$ \;
            $long\_schedule[R \ldots (R+avail] = 1$ \;
            $C_{remain} -= avail$ \;
            R = next empty slot position in long\_schedule[]\;
            \lIf {($R > D$)} {
              goto RJ
            }
     }
     $k^{++}$ \;
}
tot\_c\_iso = 0; $\Delta c_{tot} = 0$ \; \label{alg:surplus_b}
\For {each $T_i \in \mathcal{S}_{iso}$} {
 end\_BI = $\min(T_i.t_{remain\_life}, D_{max})$ \;
 tot\_c\_iso += $(T_i \cdot C_{min}) \cdot \lceil \frac{end\_BI}{T_i.P} \rceil$ \;
 $\Delta c_{tot}$ += $(T_i.C_{max} - T_i.C_{min}) \cdot \lceil \frac{end\_BI}{T_i.P} \rceil$
}
surplus\_c = $D_{max}$ - (tot\_c\_iso + tot\_async\_dur) \; \label{alg:surplus_e}
\lIf{(surplus\_c $\le$ 0)} {
 return ACCEPT, long\_schedule[]
}
\tcp*[h]{allocate surplus duration to ISO requests in a proportionally fair manner} \\
%$\Delta c_{tot} = \Sigma_{T_i \in \mathcal{S}_{iso}} (T_i.C_{max} - T_i.C_{min}) \cdot \lceil \frac{D_{max}}{T_i.P} \rceil$ \;
$\mathcal{S}_{iso}^{sort}$ = \{sorted array of all ISO requests in a non-decreasing order of their periods\} \;
\For {each $T_i \in \mathcal{S}_{iso}^{sort}$} {\label{alg:c_op_b}
  %$T_i.C_{op} = T_i.C_{min} + \min{(1, \frac{surplus\_c}{\Delta c_{tot}})} \cdot (T_i.C_{max} - T_i.C_{min})$;  \tcp{this is the $C_op$ for the entire long\_schedule[]}
  $c\_extra\_per\_job = \min{(1, \frac{surplus\_c}{\Delta c_{tot}})} \cdot (T_i.C_{max} - T_i.C_{min})$;  \tcp{this is extra duration, over $C_{min}$, available for each job of request $T_i$}

  \For {each job $J$ of $T_i$, whose release time falls within $long\_sched[]$} { \label{alg:job_starting_in_long_sched}
    allocate as much of $c\_extra\_per\_job$ to $J$ before its deadline in $long\_schedule[]$; \tcp{it may not always be possible to allocate all of $c\_extra\_per\_job$ before the job's deadline}
  }
  %$T_i.C_{op} = T_i.C_{min} + c\_extra\_per\_job$ \;
  
} \label{alg:c_op_e}
return ACCEPT, long\_schedule[] \;

\textbf{RJ:} \\
%\If {$T_n.reqType$ == ISO} \label{RJ}
\lIf {($T_n.reqType$ == ISO)} {
    $\mathcal{S}_{iso} = \mathcal{S}_{iso} - \{T_n\}$
}
\lElse {
$\mathcal{S}_{async} = \mathcal{S}_{async} - \{T_n\}$
}
return REJECT
} % of Begin

\normalsize
\end{algorithm*}

\subsection{Detailed Algorithm}
When a new request arrives to the system, the admission control algorithm called \textit{Efficient Admission Control for Isochronous and Asynchronous Requests} (EACIAR), presented in Algorithm~\ref{alg:adm_control}, is executed.
If the total utilization of all the ISO requests based on their $T_i.C_{min}$ is more than one (based on~\eqref{eq:util}, then the new request is rejected (Line~\ref{alg:iso_check}). If there is no ASYN request in the
system, then it allocates any surplus duration to the
ISO requests in a proportional fair manner (Line~\ref{alg:iso_alloc1} to \ref{alg:iso_alloc2}) and 
accepts the new request. Otherwise, it starts scheduling
from current BI till $D_{max}$, which is the maximum of current deadline of all the ASYNC requests. First,
it takes up scheduling of \textit{ISO\textsubscript{F}}  requests in the order of non-decreasing deadlines. 
For each \textit{ISO\textsubscript{F}} request $T_i$, it allocates $T_i.C_{min}$ (see Line~\ref{alg:iso_frac_alloc}).
Then, it takes up scheduling of each \textit{ISO\textsubscript{M}}  request $T_i$ in the order of non-decreasing deadlines and allocates $T_i.C_{remain}$ (see Line~\ref{alg:iso_int_alloc}). Then finally, it schedules ASYNC requests, in the non-decreasing order of their current deadlines and allocates $T_i.C_{remain}$ for each ASYNC request
$T_i$ (see Line~\ref{alg:async_alloc}). 
%At any point, if there are no time slots left, then the new request is rejected.
Then it checks for any surplus duration (over $C_{min}$) available (Line~\ref{alg:surplus_b} to Line~\ref{alg:surplus_e}). If so,
it distributes the surplus duration among the jobs of ISO requests, whose release time falls within the \textit{long\_schedule[]} interval, in a proportional fair manner (Line~\ref{alg:job_starting_in_long_sched}). It then goes back in
\textit{long\_schedule[]} and allocates as much of the surplus duration as possible,
to each job of a given ISO request before their respective deadlines (Line~\ref{alg:c_op_b} to Line~\ref{alg:c_op_e}). Note that for some ISO requests, it may not be
possible to allocate all of its share of surplus duration to all of their jobs before
their deadlines. This is because the ASYNC requests in
the system may have taken up some slots where
these surplus slots could have been allocated.
Also note that the order of handling of the type of requests is important, i.e., first the \textit{ISO\textsubscript{F}} requests should be handled followed by \textit{ISO\textsubscript{M}} requests and finally, ASYNC requests. This is to honor priority of ISO
requests over ASYNC requests\footnote{Typically ISO requests represent applications which are more critical than those represented by ASYNC requests.} and to handle ISO requests in
non-decreasing order of their respective deadlines.

When an existing request leaves the system, the same Algorithm~\ref{alg:adm_control} is used, but instead of executing it from the beginning, it starts 
executing from 
Line~\ref{alg:iso_alloc1} and continues till the end\footnote{This variation in execution between \textit{arrival of a new request} and \textit{departure of an existing request} could have been handled by distinguishing between  them at the beginning of the algorithm and bypassing relevant code for the departure case. But we chose not to clutter the algorithm with such minor details.}.

When there is at least one ASYNC request in the system, the schedule in \textit{long\_schedule[]} is followed. 
But when there is no ASYNC request in the system, EDF schedule is followed. In this case, the allocation amount
is decided as follows. 
When the system transitions from having at least one
ASYNC request to having no ASYNC request, for each \textit{ISO\textsubscript{M}} request whose $C_{remain}$ is greater than zero,
it would be allocated $C_{remain}$ by following EDF schedule. Once, its  $C_{remain}$ becomes zero, then it becomes
eligible to get allocation of $C_{op}$ in its next period.
If $C_{remain}$ of an \textit{ISO\textsubscript{M}} is less than or equal to zero, then the request becomes eligible for 
allocation of $C_{op}$ from its next period. Each
\textit{ISO\textsubscript{F}} request also becomes  eligible for
allocation of its $C_{op}$ immediately.
Note that the schedule changes only when a new request arrives or an
existing request leaves the system.

Time complexity of our EACIAR algorithm can be determined as follows. Every request goes through $BI$ 
number of time slots to compute its schedule (e.g., see
the \textit{while} loop at Line~\ref{alg:iso_frac_alloc}).
Due to the \textit{for} loop at Line~\ref{alg:for}, there are $D_{max}$  number of $BI$'s over
which each request computes its schedule. Hence,
the time complexity of our EACIAR algorithm is
$\mathcal{O}(BI \cdot D_{max} \cdot N^{iso}_{req}$), where $N^{iso}_{req}$ is the number of ISO requests in the
system.
Thus, it is a linear time algorithm.

\subsection{Correctness of the Admission Control and Scheduling}
In this section, we discuss the correctness of our EACIAR algorithm and its associated scheduling in terms of guaranteeing that no
request misses its deadline. When there is no
ASYNC request in the system, the system follows
an EDF schedule. EDF schedule guarantees that
every periodic (or ISO) request meets its deadline
since they were admitted based on \eqref{eq:util}~\cite{liu73}. When there is at least one
ASYNC request, the system follows the schedule computed in \textit{long\_schedule[]} using Algorithm~\ref{alg:adm_control}. In this algorithm, while admitting a new request, it is made sure that
the total utilization\footnote{utilization of a request $T_i$ based, on its $C_{min}$, is $\frac{T_i.C_{min}}{T_i.P}$} of the ISO requests, based on their respective $C_{min}$, is less than or equal to 1, which, as per \cite{liu73}, guarantees that the ISO requests meet their respective deadlines when EDF schedule is followed. The algorithm gives priority to ISO requests over ASYNC request while allocating
empty slots. Between \textit{ISO\textsubscript{M}} and \textit{ISO\textsubscript{F}}, \textit{ISO\textsubscript{F}} requests are picked up first (since they have shorter
deadline than \textit{ISO\textsubscript{M}}) and allocated empty slots 
equal to their respective $C_{min}$, based on EDF.
Then the \textit{ISO\textsubscript{M}} requests are scheduled in a similar manner.
Since, the utilization was calculated based on $C_{min}$ and allocation amount was also $C_{min}$, the ISO requests
would meet their respective deadlines since they follow an EDF schedule. After
scheduling all ISO requests, ASYNC requests are
scheduled based on EDF. It is made sure that each
ASYNC request meets its deadline when allocating slots to it. Finally, any surplus empty
slots are proportionally distributed to all
the ISO requests before their respective deadlines. Obviously,
this operation does not change the ability of ISO
requests to meet their deadlines. Thus, all the
requests meet their respective deadline.

%In terms of time complexity of the algorithm, it isclear that it is O(BI * D_{max} * number of 

%% file: relatedwork.tex
A few analytical channel access models for IEEE 802.11ad exist in the literature. A 3D Markov chain based model for SP and CBAP mode of channel access has been
proposed in~\cite{chen2013directional}.
\cite{pielli2020} introduces a model based on Markov chain for CBAP channel access, in the
presence of SP channel access as well as deafness and
hidden node problems due to directional antennas.
An analytical model of SP channel access is
presented in~\cite{hemanth2015} which is used to
compute worst case delay of packets. It also
studies optimal channel sharing between SP and CBAP.
An analytical model for SP channel access with
channel errors for
multimedia traffic is presented in~\cite{khorov2016}. 

However, experimental study of admission control and scheduling
of SP and CBAP channel access is quite limited in the literature. 
Authors in~\cite{lecci2021} design a max-min fair admission control and scheduling algorithm which can only handle isochronous traffic (no asynchronous traffic). They consider two very simple
application scenarios; one in which all applications have the same traffic parameter values and the other in which applications choose one
set of parameter values out of two pre-configured values. It only handles isochronous
requests having periods which are integer fraction
of a BI. It has a cubic run time complexity in
terms of number of requests.
\cite{azzino2020} presents scheduling SP channel
access using reinforcement
learning (RL). It interacts with the network
deployment scenario and uses Q-learning to 
find the optimal SP duration. The number of
error free packets received is used as \textit{reward} and the number of packets in the
MAC layer queue is used as \textit{states}.
\cite{mswim2021} proposes three admission 
control algorithms for isochronous traffic that are fair and compliant with
the IEEE 802.11ad. It also proposes an EDF based scheduler which guarantees
appropriate SP durations to the admitted isochronous
requests before their respective deadlines. These
algorithms have a linear run time complexity even
when the requests choose any arbitrary values for
their parameters and thus, is more efficient and
free of restrictions on parameters compared to~\cite{lecci2021}. Guard time overhead for 
isochronous traffic in IEEE 802.11ad MAC is accounted for in the admission control and scheduling algorithm presented in~\cite{tmc2022}. 
An admission control and scheduling algorithm 
which can handle both isochronous and asynchronous
traffic is presented in~\cite{vtc2023}. It treats asynchronous traffic as though they are periodic. This keeps the algorithm simple, but results in
overallocation of resources to asynchronous traffic and hence, leads to performance loss. In contrast,
the admission control and scheduling algorithm presented in this paper allocates only required
resources to asynchronous traffic and hence, is more efficient.

Periodic and aperiodic CPU tasks are quite similar 
to isochronous and asynchronous traffic respectively. Admission control and scheduling of
periodic and aperiodic CPU tasks have been studied in the literature~\cite{liu73, jeffay91, thuel94, chetto89}. \cite{liu73} presents admissibility of a
set of preemptive periodic task in an EDF scheduler, but it does not handle aperiodic tasks.
Schedulability conditions for non-preemptive periodic tasks are studied in~\cite{jeffay91}, but
it also does not consider aperiodic tasks. Joint scheduling of periodic and aperiodic tasks
with hard deadline is studied in~\cite{thuel94} using
a concept called \textit{slack stealing}. Basically, aperiodic tasks use the idle time or
slack time left by the periodic task. However, the
periodic task have a single parameter for execution
time, whereas in IEEE 802.11ad system, the 
isochronous requests have a range of channel time
($C_{min}$ and $C_{max}$), so the their algorithm 
cannot be applied directly. Also, in their system,
the priorities of requests are static, whereas 
in an IEEE 802.11ad system, the priorities of requests can
change as new requests arrive. \cite{chetto89} also handles scheduling of both periodic and
aperiodic requests, but using EDF which is a dynamic priority algorithm. However, their algorithm assumes that when a new aperiodic request arrives, all previously accepted aperiodic requests have finished, which is a
severe limitation. In addition, their algorithm 
requires computing exact schedule of the periodic requests in every hyper period of
the existing periodic requests\footnote{Hyper period of a set of periodic tasks is the Lowest Common Multiple of their individual periods.}. The hyper period can potentially become very large, when the periods are relatively prime. In contrast, our algorithm needs to compute the exact schedule until $D_{max}$, which is the maximum deadline among all the asynchronous requests.

%% file: conclusion.tex
We presented a comprehensive admission control algorithm, EACIAR, which handles  both isochronous and asynchronous requests in an IEEE 802.11ad MAC.
The algorithm is efficient in the sense that it
allocates resources exactly as per the 
traffic requirements of the requests and that it 
has a linear run time complexity. In addition, 
the isochronous requests get proportionally fair
allocation of SP channel time. We discussed the
correctness of the algorithm in terms of 
guaranteeing required SP duration allocation to
every admitted request before their respective 
deadlines. We believe, this is the first
comprehensive algorithm for IEEE 802.11ad MAC which handles both the traffic types and any IEEE 802.11ad compliant traffic parameters.